\documentclass{article}
\usepackage{spconf,amsmath,graphicx}

\usepackage{enumitem}
\setlist{nosep, leftmargin=14pt}
\let\svthefootnote\thefootnote
\newcommand\freefootnote[1]{%
  \let\thefootnote\relax%
  \footnotetext{#1}%
  \let\thefootnote\svthefootnote%
}
\usepackage[utf8]{inputenc}
\usepackage[english]{babel}
\usepackage{graphicx}

\usepackage{amsfonts} 
\usepackage{float}
\usepackage{xcolor}
\usepackage{caption}
\usepackage{subcaption}
\newcommand{\cmmnt}[1]{\ignorespaces}
\usepackage{eqnarray,amsmath}
\usepackage{amsmath,bm}
\usepackage{amssymb}
\usepackage{graphicx}
\usepackage[linesnumbered,ruled,vlined]{algorithm2e}

\title{MR elasticity reconstruction using statistical physical modeling and explicit data-driven denoising regularizer}
\name{%
    Narges Mohammadi$^{\star}$%$^{\star \dagger}$%
    \qquad Marvin M. Doyley$^{\star}$%
    \qquad Mujdat Cetin$^{\star}{}^{\dagger}$}
\address{$^{\star}$ Department of Electrical and Computer Engineering, University of Rochester, Rochester, NY, USA \\%
   $^{\dagger}$ Goergen Institute for Data Science, 
University of Rochester, Rochester, NY, USA
}

\begin{document}
\maketitle
\begin{abstract}
Elasticity image, visualizing the quantitative map of tissue stiffness, can be reconstructed by solving an inverse problem. Classical methods for magnetic resonance elastography (MRE) try to solve a regularized optimization problem comprising a deterministic physical model and a prior constraint as data-fidelity term and regularization term, respectively. For improving the elasticity reconstructions, appropriate prior about the underlying elasticity distribution is required which is not unique. This article proposes an infused approach for MRE reconstruction by integrating the statistical representation of the physical laws of harmonic motions and learning-based prior. For data-fidelity term, we use a statistical linear-algebraic model of equilibrium equations and for the regularizer, data-driven regularization by denoising (RED) is utilized. In the proposed optimization paradigm, the regularizer gradient is simply replaced by the residual of learned denoiser leading to time-efficient computation and convex explicit objective function. Simulation results of elasticity reconstruction verify the effectiveness of the proposed approach.
\end{abstract}

\vspace{-0.2cm}
\section{Introduction}
\vspace{-0.2cm}
%\freefootnote{This work has been partially supported by the National Science Foundation (NSF) under Grant CCF-1934962.}
%In this work, we aim to reconstruct magnetic resonance (MRE) elasticity distribution from measured harmonic displacement fields and boundary conditions (implemented as body force vector). One approach to deal with this inverse imaging problem is to obtain a linear algebraic representation for the physics-based forward model and apply it to a constrained optimization problem as a weighted least square term accompanied with a regularization term and solve it by employing proximal splitting methods.  To introduce computational benefits
\freefootnote{This work has been partially supported by the National Science Foundation (NSF) under Grants CCF-1934962 and DGE-1922591.}
MRE as an emerging elastography technique offers promising potentials for non-invasive reliable diagnosis for liver fibrosis, brain tissue degeneration, and other pathological changes using quantitative visualization of tissue properties. \cmmnt{MRE has proven successful applications for evaluating liver fibrosis, brain tissue degeneration, and other pathological changes by providing elasticity properties of each tissue} The general procedure for elasticity reconstruction of tissue can be described in two stages \cite{review2}: first, measuring deformation patterns called MRE-measurements using an MRI system in response to the external excitation through a transducer and then estimating the physical parameters of the interior medium using the measured fields \cite{MRE_book}. \cmmnt{For capturing MRE-measured data, a dynamic vibration source applied to the top exterior surface of tissue generates internal time-harmonic waves \cite{force} and the MRI system capture the displacement images with phase-contrast methods. For estimation of the tissue elasticity, different methodologies have been proposed including local frequency estimation (LFE) method, direct inversion method, and indirect model-based method using FEM \cite{sinkus2018}.} The model-based approaches for tissue elasticity estimation can be termed as a constrained optimization problem composed of physical imaging model and the prior information about the elasticity distribution which leads to improved reconstruction performance without any local homogeneity assumption (as opposed to the first two approaches)\cmmnt{ and reduced computational load by using a single frequency}. The physical imaging system describes the time-harmonic equation of motion in terms of partial differential equations (PDEs) as the forward model. Classical model-based MRE imaging approaches, employing governing PDE and physical boundary constraints, use Gaussian-Newton methods for elasticity reconstruction by assuming an initial elasticity image and solving the constrained forward model at each iteration until convergence to a stationary solution \cite{marvin} which leads to poor performance in low SNR condition \cite{newton}, \cite{MRE-AI}. Moreover, these approaches utilize fixed regularizers for various tissue patterns while appropriate data-adaptive priors might be required for capturing the complex spatial distribution of elasticity for each tissue type. Deep neural network (DNN) potentials \cite{salman2} suggests integrating the physical forward model with learning-based priors in a regularized optimization task in addition to end-to end learning applications \cite{ANN2}, \cite{mahsa1}. The integration scheme leads to both consistent reconstructions with the physical model and leveraging data-driven information using reduced amount of training pairs \cite{willet}, \cite{ali} and to accomplish this purpose, two types of approaches have been proposed: unrolling-based approaches and prior learning approaches.\\
Unrolling-based methods infuse the physical imaging model into the learning procedure by unrolling every single iteration of the optimization task as a neural network layer. This type of approaches including PINN \cite{PINN}, PI-GAN \cite{GAN}, \cite{Rezaee}, \cite{Moji}, \cite{mehdi2}, and MoDL \cite{MoDL} which perform network retraining at each optimization task iteration. On the other hand, prior learning techniques try to have learned units embedded in model-based image reconstruction by bringing learned priors as data-driven regularizers into physics-model-based image reconstruction. This group of methods including Plug-and-Play (PnP) \cite{pnp}, \cite{denoiser_prior} and regularization by denoising (RED) \cite{RED} learn a data-driven denoiser and then plug it as the regularizer proximal operator or regularizer gradient into the constrained optimization task \cmmnt{to be solved by algorithms such as GD method, alternating direction method of multipliers (ADMM) and proximal gradient approaches (PGD) \cite{primaldual}}. PnP methods substitute the proximal operator of regularizer with a DNN denoiser \cmmnt{in the constrained optimization task} which can be expressed as an implicit prior. Although PnP methods show empirical success, general theoretical convergence to the global minimum of loss function has not been provided since no explicit expression of the objective function is available. On the other hand, RED approach replaces the gradient of regularizer with the residual of denoising network leading to an explicit prior in the objective function which can provide theoretical convergence results \cite{RED2}. \\
In this paper, we propose a joint statistical and learning-based RED reconstruction \cmmnt{iterative} paradigm for estimating the MR elasticity distribution in noisy scenarios. In this methodology, the forward model of MRE imaging system is implemented as a linear algebraic representation of harmonic equilibrium equation incorporated with analytical modeling of error for elasticity distribution\cmmnt{as a united physical imaging model}. Moreover, data-driven prior information about the underlying elasticity structure of tissue type is learned using a DNN denoiser; and following RED methodology, the residual of such denoiser is plugged into the optimization task as the gradient of regularizer leading to an explicit energy function to be minimized. This underlying objective function encourages us to better understand the solution features and more properly tune the inverse problem parameters. Our simulation results using a synthetic dataset verify the improved performance of the proposed paradigm.\\
The remainder of this paper is organized as follows. We elaborate on the MRE imaging model and MRE inverse problem in Section \ref{sec2} and \ref{sec3} respectively. The proposed methodology for solving the optimization problem is introduced in Section \ref{sec4}. Our simulation and experimental results \cmmnt{of elasticity image reconstruction using synthetic MRE measurements} are provided in Section \ref{sec5}, and lastly, conclusion remarks are presented in Section \ref{sec6}. 
%synthetic wave data --> dynamic response of the structure.-->captured by harmonic oscillator model
\vspace{-0.5cm}
\section{Forward model formulation}
\vspace{-0.2cm}
\label{sec2}
\cmmnt{In the MRE imaging system, time-varying forces on the boundary of the elastic material generate a wave, or a disturbance propagates through the medium and this dynamic response of the tissue depends upon the physical properties of the tissue. }The harmonic motion equation in MRE imaging modality is governed by equilibrium constraints in terms of PDEs which reveal the relationship between the dynamic motion and tissue elasticity \cmmnt{\cite{2DMRE}}.
To simplify these governing PDEs, triangle mesh is utilized for discretization of cross-section of the medium over the mesh nodes and it is assumed that the elastic soft tissues are linear incompressible and isotropic with local homogeneity.  Leveraging the local equilibrium condition presented in \cite{narges} in each element of mesh leads to a compact linear model for the governing discretized PDEs as follows:
\vspace{-0.1cm}
\begin{equation}
\label{eq:17}
(\mathbf{k}_{e}(E)+\mathbf{k}'_{e})\mathbf{u}_{e}=\mathbf{f}_{e}
\vspace{-0.1cm}
\end{equation}
$M$ is the node numbers for each element of the mesh and $E$ represents the scalar elasticity values of the element, $\mathbf{k}_{e}(E)\in\mathbb{R}^{2M\times2M}$ is the local stiffness matrix entailing the elasticity characteristic of element, $\mathbf{k}'_{e}\in\mathbb{R}^{2M\times2M}$ containing the dynamic vibration information propagating through the element and tissue density parameter \cite{narges}, $\mathbf{u}_{e}\in\mathbb{R}^{2M\times 1}$ is the nodal Fourier deformation fields, $\mathbf{f}_{e}\in\mathbb{R}^{2M\times 1}$ is the force boundary conditions (BCs) and (\ref{eq:17}) is known as the local stiffness equation.
By considering all elements of the mesh and assembling their equivalent local equilibrium equation in the way that each nodal vector is concatenated into a global vector and each local matrix is assembled into the global one, the following global equilibrium equation can be presented by \cmmnt{\cite{narges}}:
\vspace{-0.1cm}
\begin{equation}
\begin{array}{l}
\label{eq:18}
\mathbf{K(E)}\mathbf{u}=\mathbf{D(u)}\mathbf{E}=\mathbf{f_{true}}\\
\end{array}
\vspace{-0.1cm}
\end{equation}
where $N$ indicates nodes numbers in the mesh, $\mathbf{K(E)}\in \mathbb{R}^{2N\times2N}$ containing the global stiffness information, $\mathbf{D(u)}\in \mathbb{R}^{2N\times2N}$, $\mathbf{u}\in \mathbb{R}^{2N\times1}$ is the global deformation vector, $\mathbf{E}\in \mathbb{R}^{N\times1}$ represent the elasticity distribution of the tissue over all nodes and  $\mathbf{f_{true}}\in \mathbb{R}^{2N\times1}$ denotes the Neumann BC on observed Fourier deformation vector.
\vspace{-0.2cm}
\section{Inverse Problem formulation}
\label{sec3}
The statistical representation of (\ref{eq:18}) as the forward model of MRE imaging modality unveils the relationship between the elasticity distribution $\mathbf{E}$ of tissue and the measured deformation field as follows:
\begin{equation}
\label{eq:19-1}
\mathbf{f}=\mathbf{D(u)}\mathbf{E}+\mathbf{w}\qquad \mathbf{w}\sim \mathcal{N}(0,\,\bm{\Sigma_{w}})
\vspace{-0.1cm}
\end{equation}
where $\mathbf{f}$ stands for the observed medium BCs and $\mathbf{w}\in \mathbb{R}^{2N\times 1}$ expresses the Gaussian noise. 
The frequency domain deformation measurements are acquired by applying Fourier transform to phase contrast images captured by MRI which leads to the observation process $
\mathbf{u^{m}}=\mathbf{u}+\mathbf{n}$ where $\mathbf{n}\sim \mathcal{N}(0,\,\bm{\Sigma_{n}})$ and $\mathbf{u^{m}}$ is the noisy Fourier deformation fields contaminated with noise $\mathbf{n}\in \mathbb{R}^{2N\times 1}$ with covariance $\bm{\Sigma_{n}}$.
Unifying the statistical model in (\ref{eq:19-1}) with the deformation observation process results in: 
\begin{eqnarray}
\label{eq:19-3}
\mathbf{f}&=&\mathbf{K}(\mathbf{E})\mathbf{u}+\mathbf{w}=\mathbf{K}(\mathbf{E})(\mathbf{u^{m}}-\mathbf{n})+\mathbf{w}\nonumber\\
&=&\mathbf{K}(\mathbf{E})\mathbf{u^{m}}-\mathbf{K}(\mathbf{E})\mathbf{n}+\mathbf{w}
\end{eqnarray}
Setting ${\mathbf{\Tilde{w}}}=-\mathbf{K}(\mathbf{E})\mathbf{n}+\mathbf{w}$ and employing noisy deformation fields $\mathbf{D}(\mathbf{u^{m}})\mathbf{E}=\mathbf{K}(\mathbf{E})\mathbf{u^{m}}$ yields to the following integrated observation model:
\begin{figure*}[ht]%[t]
  \centering
  \centerline{\includegraphics[width=10cm]{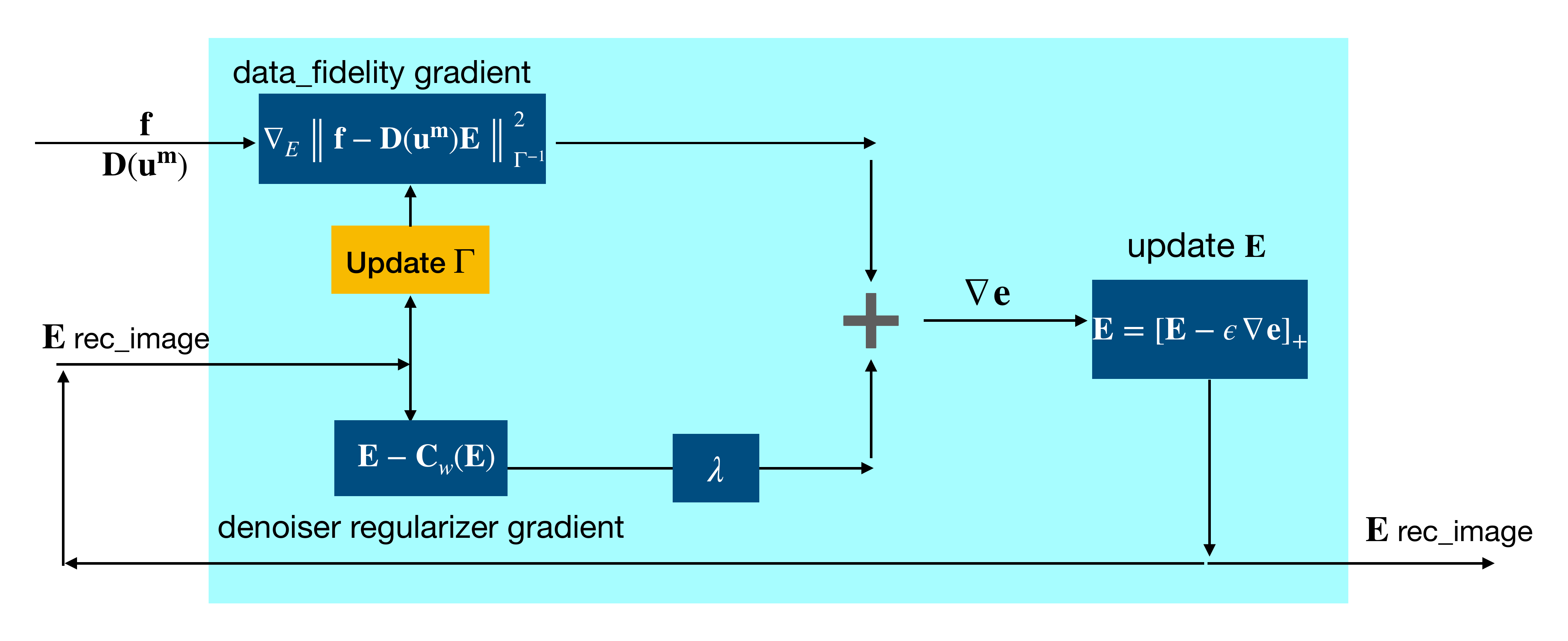}}
  \vspace{-0.45cm}
%  \centerline{(a)}\medskip
\caption{Elasticity image reconstruction using statistical physical modeling and prior learning by RED paradigm.}
\label{fig:1}
\vspace{-0.2cm}
\end{figure*}
\begin{equation}
\label{eq:19-4}
\mathbf{f}=\mathbf{D}(\mathbf{u^{m}})\mathbf{E}+\mathbf{\Tilde{w}}\qquad \mathbf{\Tilde{w}}\sim \mathcal{N}(0,\,\bm{\Gamma})
\vspace{-0.1cm}
\end{equation}
where $\bm{\Gamma}$ is expressed by:
\vspace{-0.1cm}
\begin{equation}
\label{eq:19-5}
\bm{\Gamma}=\bm{\Sigma_{w}}+\mathbf{K}(\mathbf{E})\bm{\Sigma_{n}}\mathbf{K}(\mathbf{E})^{T}
\vspace{-0.2cm}
\end{equation}
where we can mention to (\ref{eq:19-4}) as a linear forward model with signal-dependent colored noise. By acquiring $\mathbf{f}$ and $\mathbf{u^{m}}$ measurements, a regularized optimization problem has to be solved for estimating the latent elasticity distribution $\mathbf{E}$ by: 
\begin{equation}
\label{eq:20}
\begin{array}{l}
\mathbf{\hat{E}}=\mathrm{argmin} _{\mathbf{E}}\quad\frac{1}{2}\left \|  \mathbf{f}-\mathbf{D}(\mathbf{u^{m}})\mathbf{E} \right \|_{{\bm{\Gamma}}^{-1}}^{2}+\frac{N}{2}\mathrm{log}\left | \bm{\Gamma} \right |+\lambda R(\mathbf{E})\\
\quad\quad\quad
s.t.\quad \mathbf{E}>0
\end{array}
\vspace{-0.2cm}
\end{equation}
where $\left \| \mathbf{A} \right\|_{\mathbf{B}}^{2}:=(\mathbf{A}^{T}\mathbf{B}\mathbf{A})$, $R(\mathbf{E})$ is the regularization term and $\lambda$ regularization parameter. 
For solving the corresponding regularized optimization task, a fixed-point approach \cite{fixedpoint} is utilized which update $\mathbf{E}$ while $\bm{\Gamma}$ is fixed and this new $\mathbf{E}$ is employed into  (\ref{eq:19-5}) for updating $\bm{\Gamma}$. We exploit gradient descent (GD) as a first-order optimization technique to update the latent elasticity distribution $\mathbf{E}$ as follows:
\vspace{-0.2cm}
\begin{equation}
\label{eq:20-1}
\mathbf{E}\xleftarrow{}[\mathbf{E}-\epsilon (\nabla g({\mathbf{E}})+\lambda \nabla R({\mathbf{E}}))]_{+}
\vspace{-0.3cm}
\end{equation}
\vspace{-0.3cm}
where:
\begin{equation}
\label{eq:23}
g(\mathbf{E})=\frac{1}{2}(\mathbf{f}-\mathbf{D}(\mathbf{u^{m}})\mathbf{E}  )^{T}\bm{\Gamma}^{-1}(\mathbf{f}-\mathbf{D}(\mathbf{u^{m}})\mathbf{E}  )
\end{equation}
\begin{equation}
\label{eq:24}
\nabla g(\mathbf{E})=-(\mathbf{D}(\mathbf{u^{m}}))^{T}\bm{\Gamma}^{-1}(\mathbf{f}-\mathbf{D}(\mathbf{u^{m}})\mathbf{E} )
\end{equation}
where $\epsilon$ is the step-size,  $[]_{+}$ holds for the positivity constraint on elasticity reconstruction and $\nabla R({\mathbf{E}})$ is introduced in the next Section.
\vspace{-0.2cm}
\section{prior learning by RED methodology}
\label{sec4}
\vspace{-0.1cm}
For reconstructing the elasticity distribution $\mathbf{E}$, the prior information about the underlying pattern of latent elasticity images and its equivalent gradient should be incorporated in (\ref{eq:20}) and (\ref{eq:20-1}). Here, we explore the potential of data-adaptive regularizers to capture the complex spatially varying patterns of elasticity distribution. RED methodology introduces a computation-efficient approach for solving the elasticity inverse problem which consists of supervised learning of a denoiser network and applying the residual of learned denoiser as the regularizer gradient in (\ref{eq:20-1}).
Following RED paradigm, the gradient of the regularization term is expressed as:
\begin{equation}
\label{eq:25}
\nabla R(\mathbf{E})=\mathbf{E}-\mathbf{C}_{w}(\mathbf{E})
\end{equation}
where $\mathbf{C}_{w}(.)$ is a learned denoiser network parameterized with weights $w$ which satisfies RED conditions \cite{RED}, \cite{RED2}. It is worth mentioning that in (\ref{eq:25}) no gradient computation is performed and instead, simple residual of denoiser is employed as the gradient which highlights the RED power for reducing the computational costs. Moreover, an explicit expression for the regulaizer and consequently the objective function can be presented which enables the convergence analysis and more efficiently parameter tuning including $\lambda$. To this regard, the RED explicit regularization term can be described as follows:
\vspace{-0.3cm}
\begin{equation}
\label{eq:21}
R(\mathbf{E})_{RED}=\frac{1}{2}\left \langle \mathbf{E},\mathbf{E}-\mathbf{C}_{w}(\mathbf{E}) \right \rangle=\frac{1}{2}\mathbf{E}^{T}(\mathbf{E}-\mathbf{C}_{w}(\mathbf{E}))
\end{equation}
which introduce the elasticity inverse problem in (\ref{eq:20}) as a convex optimization problem if the denoiser network satisfies contractivity condition \cite{RED2}.
\cmmnt{Benefiting the explicit objective function introduced by (\ref{eq:20}) and (\ref{eq:21}), the regularization parameter $\lambda$ can be tuned efficiently using the noise level information. In this regard, we utilize strong passivity condition of RED \cite{RED} which indicates that $\left \| \nabla _{\mathbf{E}} \mathbf{C}_{w}(\mathbf{E}) \right \|\leq 1$, thus, $\lambda$ at the equilibrium point can be estimated by:
\begin{equation}
\label{eq:26}
\lambda \geq \left \| \nabla g(\mathbf{E}) \right \|
\end{equation}
}
It should be mentioned that the denoising network $\mathbf{C}_{w}$ is trained using ground-truth elasticity images and poor noisy ones $\tilde{\mathbf{E}}$ generated as \textit{maximum likelihood} (ML) estimation of elasticity by solving the unregularized form of the optimization problem in (\ref{eq:20}) using mean squared error (MSE) cost function as follows:
\vspace{-0.2cm}
\begin{equation}
\label{eq:22}
\emph{l}(w)=\frac{1}{2N}\sum^{N}\left \| C_{w}(\tilde{\mathbf{E}})-\mathbf{E} \right \|_{F}^{2}
\end{equation}
Once $\mathbf{C}_{w}$ is trained its residual is plugged into the iterative estimation scheme in (\ref{eq:20-1}).\\
\begin{figure*}[t!]%[htb]
\begin{minipage}[b]{0.19\linewidth}
  \centering
  \centerline{\includegraphics[width=3.0cm]{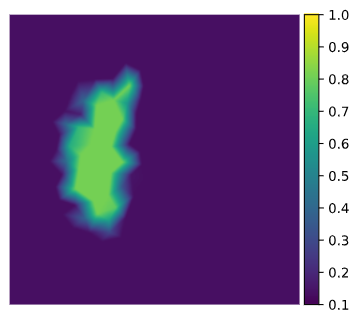}}
    \vspace{-0.6\baselineskip}
%  \vspace{1.5cm}
  \centerline{ \scriptsize{}}\medskip
\end{minipage}
\begin{minipage}[b]{.19\linewidth}
  \centering
  \centerline{\includegraphics[width=2.4cm]{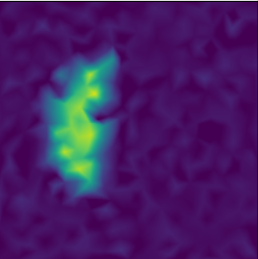}}
    \vspace{-0.6\baselineskip}
%  \vspace{1.5cm}
  \centerline{ \scriptsize{}}\medskip
\end{minipage}
%\hspace{1cm}
\begin{minipage}[b]{0.19\linewidth}
  \centering
  \centerline{\includegraphics[width=2.4cm]{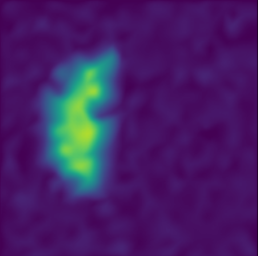}}
    \vspace{-0.6\baselineskip}
%  \vspace{1.5cm}
  \centerline{ \scriptsize{}}\medskip
\end{minipage}
%\par\vspace{+0.25\baselineskip}
\begin{minipage}[b]{0.19\linewidth}
  \centering
  \centerline{\includegraphics[width=2.4cm]{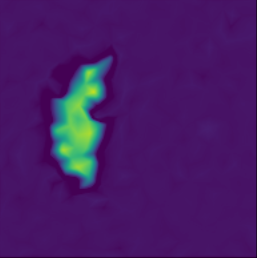}}
    \vspace{-0.6\baselineskip}
%  \vspace{1.5cm}
  \centerline{ \scriptsize{}}\medskip
\end{minipage}
%\hspace{1cm}
\begin{minipage}[b]{0.19\linewidth}
  \centering
  \centerline{\includegraphics[width=2.4cm]{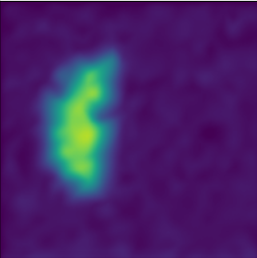}}
    \vspace{-0.6\baselineskip}
%  \vspace{1.5cm}
  \centerline{ \scriptsize{}}\medskip
\end{minipage}
%\hspace{1cm}
\par\vspace{-0.5\baselineskip}
\begin{minipage}[b]{0.19\linewidth}
  \centering
  \centerline{\includegraphics[width=3.0cm]{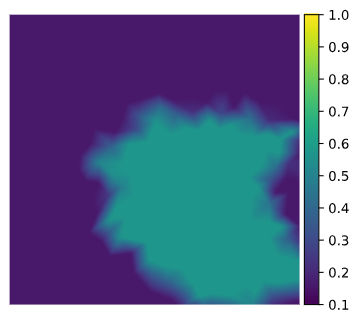}}
    \vspace{-0.5\baselineskip}
%  \vspace{1.5cm}
  \centerline{(a) \scriptsize{}}\medskip
\end{minipage}
\begin{minipage}[b]{.19\linewidth}
  \centering
  \centerline{\includegraphics[width=2.4cm]{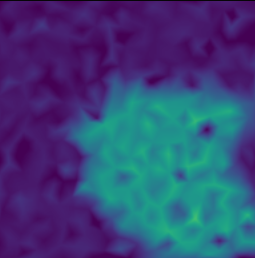}}
    \vspace{-0.3\baselineskip}
  %\vspace{1.5cm}
  \centerline{(b) \scriptsize{}}\medskip
\end{minipage}
%\hspace{1cm}
\begin{minipage}[b]{0.19\linewidth}
  \centering
  \centerline{\includegraphics[width=2.4cm]{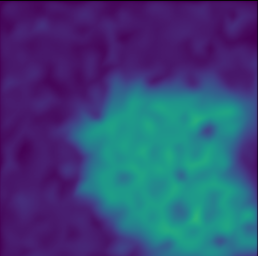}}
    \vspace{-0.3\baselineskip}
%  \vspace{1.5cm}
  \centerline{(c) \scriptsize{}}\medskip
\end{minipage}
%\par\vspace{+0.25\baselineskip}
\begin{minipage}[b]{0.19\linewidth}
  \centering
  \centerline{\includegraphics[width=2.4cm]{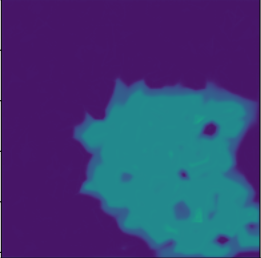}}
    \vspace{-0.3\baselineskip}
%  \vspace{1.5cm}
  \centerline{(d) \scriptsize{}}\medskip
\end{minipage}
%\hspace{1cm}
\begin{minipage}[b]{0.19\linewidth}
  \centering
  \centerline{\includegraphics[width=2.4cm]{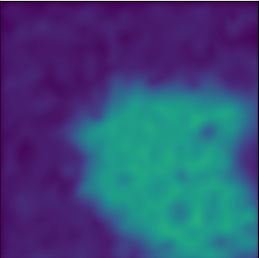}}
    \vspace{-0.3\baselineskip}
%  \vspace{1.5cm}
  \centerline{(e) \scriptsize{}}\medskip
\end{minipage}
%\hspace{1cm}
\vspace{-0.35cm}
\caption{(a) Ground-truth elasticity image. (b) Reconstructed elasticity image using unregularized optimization. (c) Reconstructed image with post-processing approach. (d) Reconstructed image using PnP approach and DnCNN. (e) Reconstructed elasticity image with the proposed denoising regularizer by RED approach. The unit of the color bar is 100 KPa. }
\vspace{-0.5cm}
\label{fig:3}
\end{figure*}
\begin{figure}[h]%[htb]
\begin{minipage}[b]{0.8\linewidth}
  \centering
  \centerline{\includegraphics[width=5.8cm]{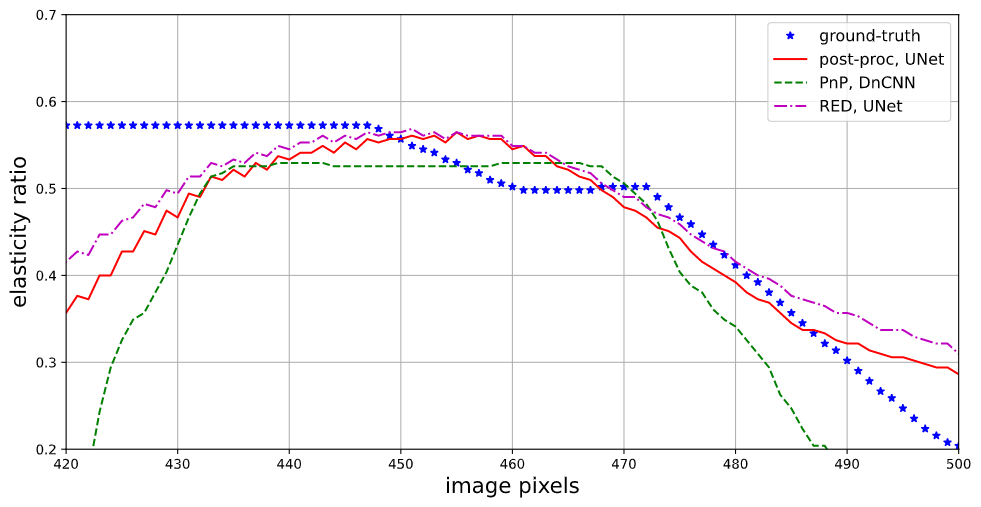}}
    \vspace{-0.4\baselineskip}
%  \vspace{1.5cm}
  \centerline{(a) \scriptsize{}}\medskip
\end{minipage}
\begin{minipage}[b]{.19\linewidth}
  \centering
  \centerline{\includegraphics[width=1.7cm]{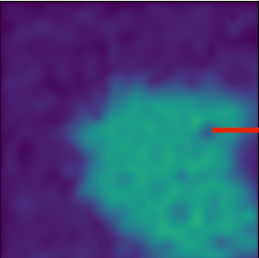}}
    \vspace{+0.4\baselineskip}
%  \vspace{1.5cm}
  \centerline{(b)\scriptsize{}}\medskip
\end{minipage}
%\hspace{1cm}
  \vspace{-0.85cm}
\caption{ (a) The cross section of reconstructed elasticity images using different approaches marked by the red line in (b). }
\label{fig:4}
\vspace{-0.55cm}
\end{figure}
The overall reconstruction procedure is depicted in Fig. \ref{fig:1}. The initial elasticity distribution $\mathbf{E}$ in this methodology is a poor ML estimate of elasticity image. RED approach as a prior learning paradigm incorporates the separate roles of the data-fidelity term and a data-driven regularization term. Regarding the data-fidelity term which represents the statistical physical imaging model and following the fixed-point technique, covariance matrix $\Gamma$ is updated using the current estimate of $\mathbf{E}$ and then this fixed $\Gamma$ is employed in the physical forward model for computing the data-fidelity gradient. On the other hand, the gradient of the denoiser regularizer is easily obtained by the residual of the denoiser network; and finally, the gradient of both data-fidelity and regularizer terms are incorporated in the GD update iterates. It is noteworthy to mention that as the statistical physical model is derived for the imaging system, no massive network parameters are required for learning the physical model and this fact significantly reduces the required training dataset size.
\vspace{-0.2cm}
\section{Simulations and Results}
\label{sec5}
For performance evaluation of the proposed method, we seek to reconstruct the elasticity distribution image $\mathbf{E}$ using the measured noisy Fourier deformations $\mathbf{u^{m}}$ (also called phase difference fields) and Neumann BCs presented as  $\mathbf{f}$. The denoiser network $\mathbf{C}_{w}$ needs to be trained using true elasticity images and poor noisy ones acquired by solving the unregularized optimization problem in (\ref{eq:20}). In this regard, a dataset of 541 mask images \cite{dataset} of lesions embedded in background tissues is utilized for generating ground-truth elasticity images (synthetic maps). We generate normalized elasticity for each lesion in the range of 0.3-0.8 KPa and normalized elasticity for background tissue in the range of 0.1-0.15 KPa. With respect to these settings, the lesion elasticity to the background elasticity ratio remains in the range of 2-8 which expresses experimental conditions well. The medium cross-section image is discretized using triangle elements resulting in MRE measurements with $2N\times1$ dimension expressing the axial and lateral measurements over the mesh nodes. Utilizing this generated dataset, the deformation fields $\mathbf{u^{m}}$ are obtained by solving the deterministic forward model $\mathbf{K(E)}\mathbf{u}-\mathbf{f}=0$ and adding multivariate Gaussian noise \cite{salman1} $\mathbf{n}$ with $SNR=35dB$ \cmmnt{for each synthetic true elasticity image with known elasticity distribution $\mathbf{E}$}. Other setting in MRE imaging including the tissue density parameter for soft tissues which are mainly made of water is set to $\rho=1000Kg/m^{3}$ and the transducer excitation frequency is $\omega=90Hz$ (which have been used for constructing $\mathbf{k}'_{e}$ in (\ref{eq:17})).
For training the denoiser network, training pairs including noisy elasticity images by solving unregularized optimization task and ground-truth elasticity images of size $512\times512$ are fed into the UNet architecture. \cmmnt{  For generating noisy elasticity images, we map the noisy displacement fields $\mathbf{u^{m}}$ to the image domain by solving the unregularized inverse problem consisting of a data-fidelity term and a positivity constraint.}
To evaluate the reconstruction performance, two other supervised learning paradigms are implemented including the post-processing approach \cite{postprocessing} deployed by UNet architecture and PnP method using DnCNN structure \cite{narges2}, \cite{tum}. For all implemented approaches, the batch-size=16, epochs-number=100, learning rate of Adam optimizer is lr=1e-3 \cmmnt{and other network architecture details are presented in Table \ref{table1}}.\\
The elasticity images reconstructed using these methods are illustrated in Fig. \ref{fig:3}. Both the post-processing and RED approaches using UNet architecture try to remove the artifacts while making the image blurred and the advantage of RED over the post-processing approach is that the strength of denoiser can be controlled by the regularizer parameter. The PnP approach using DnCNN is able to effectively preserve the edges while is less efficient in reducing the artifacts concerning the internal texture. More detailed comparison of reconstructed images is presented in Fig. \ref{fig:4} in terms of cross-section values of elasticity modulus $\mathbf{E}$ which demonstrates the effectiveness of the RED approach by less error with respect to the ground-truth one. Regarding computation time, iterative schemes of the PnP and RED approach for updating the reconstruction leads to more computation time in comparison with the post-processing method. Moreover, the PnP approach uses proximal gradient methodology which is slower than gradient descent used in the proposed RED approach.     
\cmmnt{
\begin{table}[h]
\centering\begin{tabular}{ | m{8em} | m{5cm}| } %| m{0.6cm} 
\hline
method & architecture \\ 
\hline
Post-processing and RED reconstruction & UNet: 4 strided conv. (stride=2) + 4 transposed conv., Leaky ReLU and skip connection after each layer \\
\hline
PnP reconstruction & DnCNN : 1 strided conv. (stride=1) + 10 (conv. layer,BN) + 1 conv., ReLU after each layer except the last one \\ 
\hline
\end{tabular}
\vspace{-0.25cm}
\caption{Details of implemented networks for MR elasticity reconstruction.}
\label{table1}
\vspace{-0.5cm}
\end{table}
}
\vspace{-0.3cm}
\section{Conclusion}
\vspace{-0.2cm}
\label{sec6}
In this article, we proposed a statistical and data-driven approach for FEM-based MR elastography by solving a constrained optimization problem. This methodology presents an explicit joint objective function composed of a statistical forward model \cmmnt{representing the MRE harmonic motion equations} and a learning-based regularizer for capturing the underlying prior information about the elasticity pattern. The statistical physical model introduces a linear representation with respect to the latent elasticity distribution and a signal-dependent colored noise model. A denoising network is supervisedly trained and its residual is simply replaced as the regularizer gradient which leads to an explicit representation of the objective function and reduced computation time. \cmmnt{Leveraging theoretical convergence and uniqueness guarantees of RED approach, the corresponding optimization task is solved by fixed-point iterative paradigm and GD which leads to more stale elasticity estimation.} Since the proposed scheme is a root for 3D MRE image reconstruction, ease of deploying the regularizer gradient is a significant gain. The simulation and comparison results verify the robustness and effectiveness of the proposed paradigm. 
\vspace{-0.3cm}
\bibliographystyle{IEEEbib}
\bibliography{M335}
%\printbibliography

\end{document}